\documentclass{article}
\usepackage{ijcai23}

\usepackage{times}
\usepackage{soul}
\usepackage{url}
\usepackage[hidelinks]{hyperref}
\usepackage[utf8]{inputenc}
\usepackage[small]{caption}
\usepackage{graphicx}
\usepackage{amsmath}
\usepackage{amsthm}
\usepackage{booktabs}
\usepackage{algorithm}
\usepackage{algorithmic}
\usepackage[switch]{lineno}

\title{A Survey for Graphic Design Intelligence}

\author{
Danqing Huang$^{1\dagger}$\and
Jiaqi Guo$^{1\dagger}$\and
Shizhao Sun$^{1\dagger}$\and
Hanling Tian$^{2\mathsection}$\and
Jieru Lin$^{3\mathsection}$\and
Zheng Hu$^{4\mathsection}$\and
Chin-Yew Lin$^{1}$\and
Jian-Guang Lou$^{1}$\and
Dongmei Zhang$^{1}$
\thanks{$^\dagger$Equal contributions. $^\mathsection$This work was done when the authors were interns at Microsoft Research Asia.}
\affiliations
$^1$Microsoft 
$^2$Xi'an Jiaotong University \\
$^3$Harbin Institute of Technology
$^4$ Central South University\\
\emails
\{dahua,shizsu,jiaqiguo,cyl,jlou,dongmeiz\}@microsoft.com,
thl2628741788@stu.xjtu.edu.cn,
hitjierulin@gmail.com,
huzhengwuyu@gmail.com
}

\begin{document}

\maketitle

\begin{abstract}
Graphic design is an effective language for visual communication. Using complex composition of visual elements (e.g., shape, color, font) guided by design principles and aesthetics, design helps produce more visually-appealing content. The creation of a harmonious design requires carefully selecting and combining different visual elements, which can be challenging and time-consuming. To expedite the design process, emerging AI techniques have been proposed to automatize tedious tasks and facilitate human creativity. However, most current works only focus on specific tasks targeting at different scenarios without a high-level abstraction. This paper aims to provide a systematic overview of graphic design intelligence and summarize literature in the taxonomy of representation, understanding and generation. Specifically we consider related works for individual visual elements as well as the overall design composition. Furthermore, we highlight some of the potential directions for future explorations.
\end{abstract}

\section{Introduction}

Graphic design conveys messages in a visual way.
It occurs everywhere in our daily life, from striking advertisements, attractive movie posters, to branding logos and stunning user interfaces.
An aesthetically-pleasing design can draw audiences' attention and improve their engagement.

\begin{figure}[tb]
\centering
\includegraphics[width=0.9\linewidth]{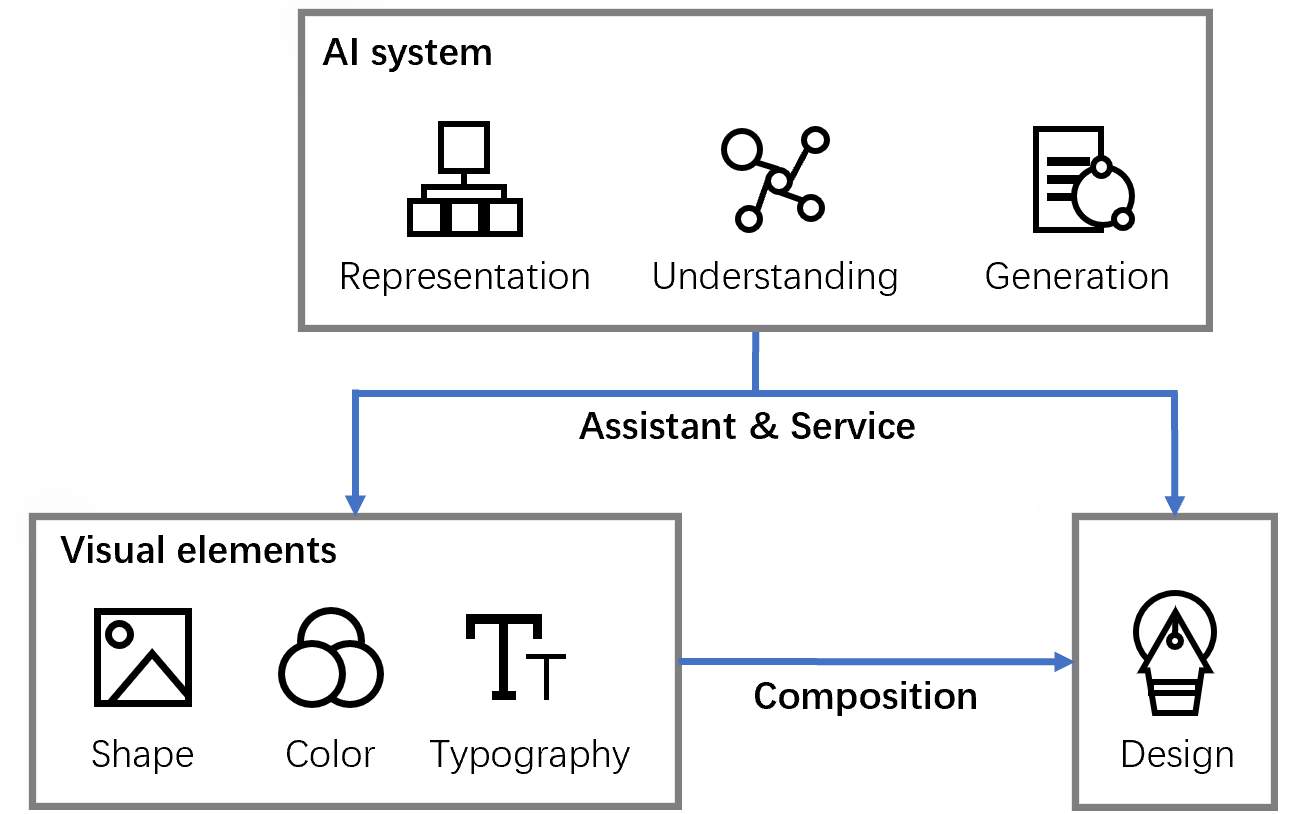}
    \caption{An overview of graphic design intelligence. The composition of visual elements produces a design. AI components including representation, understanding and generation serve as assistants to enhance graphic design in both element-level and design-level.}
    \label{fig:overview}
\end{figure}

However, creating a successful design can be difficult and time-consuming, due to the enormous space of design choices.
For instance, to create a product advertising poster, a designer often first explores posters for similar products to form an initial idea. Then, she needs to select or create visual elements (e.g., shape, color, font \& typography) that best match the product and the poster style. 
Finally, all the visual elements are composited properly, such that the resulting poster is compelling and conforms to the well-developed design practices (e.g., the golden ratio principle).
The above process usually iterates multiple times to finalize the poster.
To ease this creation process, practitioners and researchers have innovated a lot of professional authoring tools, such as Photoshop, Illustrator, and Sketch, but it still requires huge human efforts to create designs.

Recently, with the rapid advancements in artificial intelligence (AI), the field of \textbf{Graphic Design Intelligence} has been emerging, which applies AI to innovating intelligent services for graphic design (see Figure~\ref{fig:overview}).
Research work in this field can be mainly categorized into three groups.
The first line of work, \textit{Graphic Design Representation}, aims to learn low-dimensional and semantic-rich representations for designs.
The representations make it possible to search for designs with targeted functionalities and styles, accelerating the ideation process.
The second line of work, \textit{Graphic Design Understanding}, attempts to perform reverse engineering on certain aspects of designs, such as identifying visual elements, parsing element relations, and recognizing visual flows.
The design understanding capability empowers a rich set of applications, e.g., visual element recommendations, design quality assessments, and focus area prediction.
The third line of work, \textit{Graphic Design Generation}, takes a step further and aims to automate the complicated design creation process.
Given a specification, it automatically generates visual elements and composites them to form a compelling design.
Nowadays, as the techniques developed, an increasing number of intelligent services have been integrated into professional tools to boost productivity, and it also raises an urgent need to comprehensively organize existing literature for a better overview of graphic design intelligence.

In this survey, we aim to bridge this gap and review the related literature on graphic design intelligence according to the taxonomy of representation, understanding, and generation.
We also discuss the open questions in this field and some promising research directions.

\section{Overview}

In this section, we introduce some graphic design basics and summarize the major components in design intelligence.

\subsection{Graphic Design Basics}
\label{subsec:basic}

Graphic design is the process of planning and projecting ideas with the use of rich visual elements. It helps present the content in a more visually-appealing way.
There are many types of design such as publication (book, magazine), user interface (website, mobile app) and advertising (poster, banner). As it covers a wide range of fields, this survey focuses on key and general AI techniques applied in static graphic design under digital media.

\paragraph{Visual Elements} are the building blocks of graphic design. Here we list some most frequently-used elements:
\begin{itemize}
    \item \textbf{Shape.} It is recognizable objects defined by boundaries such as with lines, color, or negative space. Shape types are diverse which can be flat (2D) or solid (3D), geometric or organic, etc.
    \item \textbf{Color.} It carries meanings such as theme, mood and visual hierarchy using attributes such as hue, saturation, value and brightness. 
    \item \textbf{Font \& Typography.} Fonts with different leading, tracking and kerning can represent different meanings. Typography arranges type in order to make the displayed words legible and appealing. 
\end{itemize}

\paragraph{Layout and Composition.} Together considering the individual visual elements, we use principles in layout and composition (e.g., alignment, contrast, repetition) to arrange elements and create a harmonious design. 

\subsection{AI in Graphic Design}
AI technologies have successfully empowered many applications in graphic design, facilitating human to speed up the design process including materials searching, ideation, design creation and refinement, etc. 
From the view of machine learning, graphic design intelligence can be summarized into three core components:
\begin{itemize}
    \item\textbf{Representation} relates to how to encode a design which serves as a backbone to surface tasks. It is an important factor for the success of applying machine learning algorithms. (\textbf{Section~\ref{sec:repr}})
    
    \item\textbf{Understanding} focuses on machine comprehension to analyze the syntax and semantic of a design. It aims to extract the hidden information (e.g. logical structure, style) from the unstructured design. (\textbf{Section~\ref{sec:understand}})
    
    \item\textbf{Generation} is the process of producing novel visual response (e.g., layouts, colors, fonts) based on input constraints or queries, which improves design creativity efficiency. (\textbf{Section~\ref{sec:gen}})
\end{itemize}

As shown in Figure~\ref{fig:overview}, each of the three components enhances graphic design in both visual-element-level and overall design-level. In the following sections, we will introduce each of them in more details.


\section{Representation}
\label{sec:repr}

Representation learning aims to train machine-learning algorithms for learning useful representations of data, such that building classifiers and predictors becomes easier and more efficient~\cite{Bengio2013Representation}.
In this section, we review the literature on learning element-level representations and design-level representations.

\subsection{Element-level Representation}
\label{subsec:repr-element}
This line of work aims to model the unique characteristics of visual elements and learn their representations accordingly.

\paragraph{Shape.}
Shapes in graphic design are typically represented as vector graphics, which can be rendered at arbitrary resolutions (i.e., scale-invariant).
To learn a shape embedding, Lopes~\shortcite{Lopes2019Learned} proposes the first neural model.
It consists of a variational autoencoder (VAE)~\cite{kingma2013auto} that learns to project a rasterized shape into a latent space and an auto-regressive decoder that generates a shape by predicting a sequence of discrete commands (e.g., \texttt{moveTo}).
The VAE and decoder are optimized independently.
Albeit effective, this approach requires clean discrete commands as labels, which are difficult to obtain in practice.
To mitigate this limitation, Reddy~\shortcite{Reddy2021Im2Vec} presents Im2Vec, a VAE-based approach that learns shape embeddings with indirect supervision from rasterized shapes.
The encoder in Im2Vec takes a rasterized shape as input and generates a latent code.
A set of decoders then predict Bezier paths based on the code to reconstruct the shape.
The paths can be rasterized and composited differentiablely so that Im2Vec can be optimized in an end-to-end manner without any path supervision.
The learned shape embeddings enable plausible interpolation between samples to create novel shapes.

\paragraph{Color.}
Some research efforts have attempted to learn neural representations of color for palette recommendation and re-colorization tasks.
Kim~\shortcite{Kim2022Colorbo} regards each color as a discrete token and a color palette as a sentence.
They adopt the fastText embedding technique~\cite{bojanowski-etal-2017-enriching}, 
to obtain color embeddings.
Qiu~\shortcite{qiu2022color} takes a step forward and attempts to learn contextualized color embeddings using Transformer~\cite{Transformer} and BERT-style training loss~\cite{BERT}.
Unlike this line of work, Deshpande~\shortcite{Deshpande_2017_CVPR} aims to learn a disentangled latent color representation for whole images or designs using VAE.
The color representations can then be used to colorize grayscale images.

\paragraph{Font \& Typography.}
Early work~\cite{Wang2015DeepFont} on typography representations ratsterizes text of a specific font into an image and leverages the Convolutional neural network (CNN) to obtain the font embedding.
The CNN is trained with the image reconstruction loss and the font family classification loss.
Later work~\cite{Balashova2018LearningAS} exploits the structural nature of typography and attempts to learn stroke-based font representations using approaches similar to those for vector graphics.
Liu~\shortcite{Liu2022Learning} instead proposes to model a glyph as shape primitives enclosed by quadratic curves.
They find this implicit representation well-suited because glyphs have clear boundaries and can be easily split into parts.
Reddy~\shortcite{Reddy2021Multi} extends this idea further and develops the multi-implicit representation for glyphs based on the observations that a complex font can be encoded as a composition of global implicit functions.
The learned font representations can be used for font retrieval, style transfer, and novel typography creation.

\subsection{Design-level Representation}
\label{subsec:repr-design}

This spectrum of work seeks to learn representations of a design based on how it is composited with visual elements. 

\paragraph{Layout.}
Layout, the positioning and sizing of all the shapes in a design, plays an important role in graphic design. 
Deka~\shortcite{RICO} renders a layout as an image and learns a low-dimensional layout embedding with an autoencoder.
To model the geometric relationships between shapes more effectively, Manandhar~\shortcite{Manandhar2020Learning} proposes to represent a layout as a graph, in which vertices indicate shapes and edges denote the geometric relationships.
A graph neural network is adopted to encode the layout graph and optimized via the layout reconstruction loss and a triplet ranking loss.
Xie~\shortcite{Xie2021CanvasEmb} introduces CanvasEmb, which takes a richer set of shape properties into consideration for layout modeling, including type property, geometric property, and content-related properties.
The Transformer encoder is employed to encode a layout and is trained with the masked language modeling objective~\cite{BERT}.
The learned layout representations can facilitate various layout understanding tasks, e.g., layout retrievals, object role detection, and relation parsing~\cite{feng2022rep}.

\paragraph{Overall Design.} 
A design is often composed of visual elements with multi-modal content.
For instance, a graphic user interface (GUI) includes textual and visual content.
Hence, it is natural to explicitly model this compositionality with modality-aware feature extractors.
He~\shortcite{He2021ActionBert} proposes ActionBERT to learn mobile GUI representations.
Given a GUI, ActionBERT first tries to detect all its visual elements through the view hierarchy or by using OCR and object detection services.
Then, ActionBERT leverages a pre-trained text encoder and vision encoder to extract features for elements with multi-modal content.
The features are finally fed into a Transformer encoder to build the GUI representation.
Rastogi~\shortcite{UIBert} introduces UIBert, which differs from ActionBERT in two aspects.
First, UIBert leverages a richer set of features from the GUI view hierarchy.
Second, UIBert primarily focuses on representing a single GUI, while ActionBERT supports multiple GUIs as input, enabling multi-GUI downstream tasks.
Fu~\shortcite{fu2021understanding} targets the same venue as UIBert and introduces PW2SS.
In addition to the multi-modal content features, PW2SS incorporates a layout embedding module to extract global layout information.
The learned GUI representation can facilitate a wide range of downstream tasks, e.g., UI retrieval, application classification, icon classification, etc.

While explicitly modeling a design based on its compositionality is sensible and effective, it requires access to its internal structure via view hierarchy or a set of off-the-shelf services, which are not easily accessible and could be error-prone.
To alleviate this shortcoming, some recent work attempts to treat a design as an image and implicitly learn its compositionality via a well-designed pre-training task.
Specifically, Lee~\shortcite{lee2022pix2struct} presents Pix2Struct, an image-to-text model for design representation.
Given a design, Pix2Struct rasterizes it as an image and obtains its representation with a vision encoder.
To acquire the compositionality knowledge, they propose a screenshot parsing objective to pre-train the model, which is to predict the serialized structure of a given design, including its element type, hierarchy, and textual content.
To facilitate the pre-training, they collect an 80-million web page dataset.
Though pre-trained on Web GUI, Pix2Struct can transfer its knowledge to a wide variety of designs, such as mobile GUI, book covers, and infographics, and it presents promising downstream performance.

\subsection{Discussion}
\label{subsec:repr-discussion}

From the literature review above, one can observe the connection between element-level work and design-level work is quite loose. 
This is unexpected because a design is intentionally composited by its visual elements.
The representations of visual elements are supposed to benefit design representations, and vice versa.
We believe that enhancing this connection deserves more careful study.
In addition, an unlimited number of graphic designs are available on the internet, and they are treasured resources for building general-purpose design representation models.
However, these designs do not have precise metadata, such as view hierarchy, shape position and size, causing a huge challenge for learning representations.
Pix2Struct~\cite{lee2022pix2struct} makes a pioneer attempt in this direction by discarding metadata, and there is still a long way to go.

\section{Understanding}
\label{sec:understand}

Understanding aims to perform reverse engineering from an unstructured design and extract useful information such as style and visual flow.
In this section, we introduce several key aspects of design understanding in both element-level and design-level.
We show an example in Figure~\ref{fig:understanding}.

\subsection{Element-level Understanding}
Here we summarize the understanding tasks and methods in the context of individual visual elements.

\begin{figure}[htb]
\centering
\includegraphics[width=0.75\linewidth]{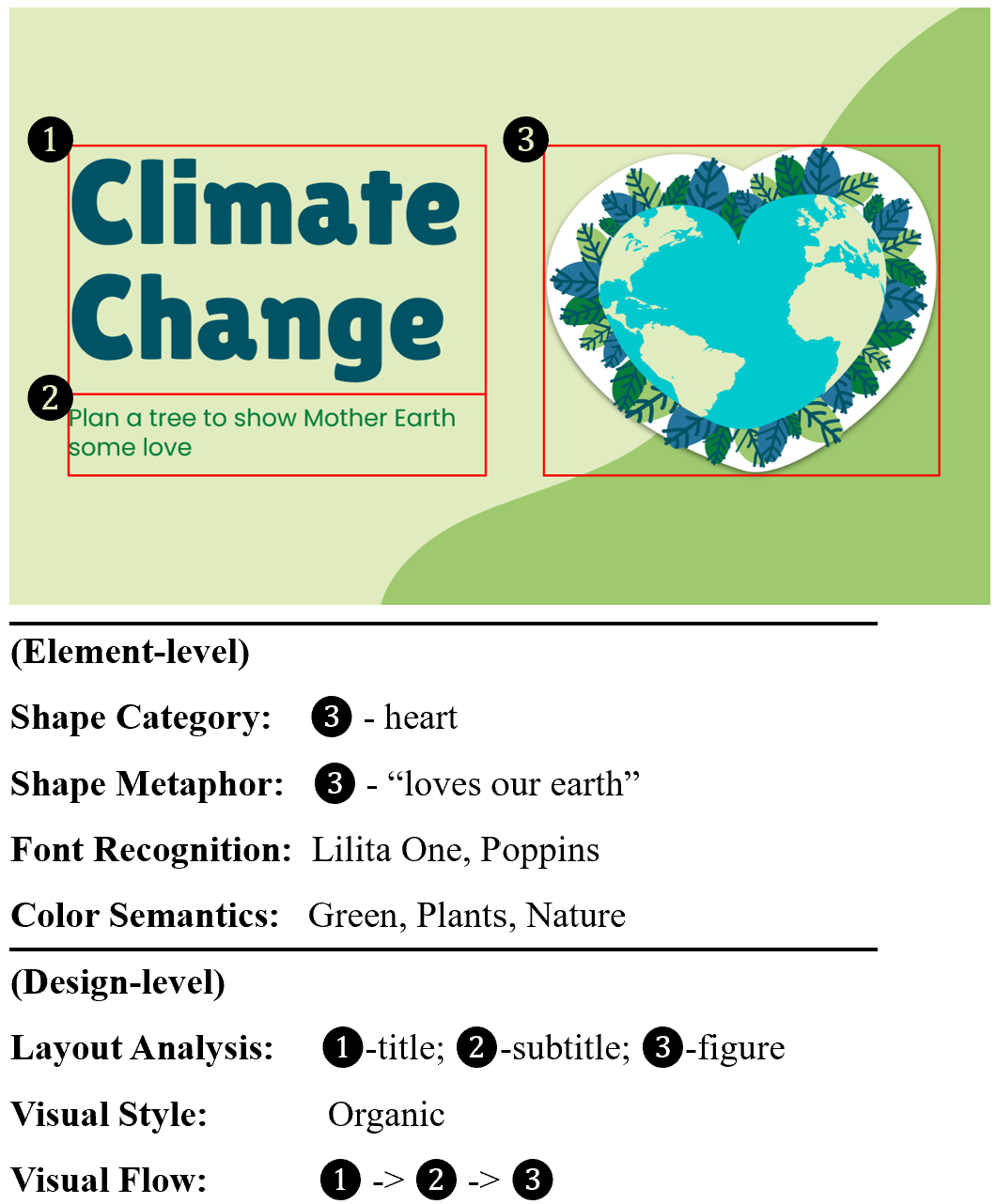}
    \caption{Design understanding in element-level and design-level.}
    \label{fig:understanding}
\end{figure}

\paragraph{Shape.} Using arbitrary curves, shapes are attached with meanings: (1) \textbf{Explicit objects:} shapes can represent real-world objects (e.g., plane, apple). A series of works have focused on sketched shape recognition, which classify hand-drawn shapes to object categories. Recent approaches have explored different backbone architectures such as CNN, RNN and Transformer~\cite{jongejan2016quickdraw,lin2020sketch}) for sketch sequences encoding and classification.
(2) \textbf{Implicit metaphors:} visual metaphor is a creative way to deliver concepts by implying through visual symbols and juxtaposition, which is widely used in advertising. As an example shown in Figure~\ref{fig:understanding}, the heart-like shape infers ``\textit{loves our earth}''.
Theoretical investigations with user studies~\cite{Indurkhya2017Interpreting} have been conducted to understand visual metaphors.
Creating and interpreting visual metaphors is challenging, which requires mappings between visual objects and common knowledge.
Recently there are ideation tools to facilitate visual metaphors creation.
DARCI~\cite{derrall2014metaphor} learns the relatedness between adjectives (e.g., dark, cold) and low-level image features with neural networks. MetaMap~\cite{kang2021metamap} explores relevant metaphor examples with extracted semantic features (concept graph), color and shape for metaphor recommendations.

\paragraph{Color.} 
Progress has been made in exploring color semantics. \textit{Color Image Scale}~\cite{kobayashi1991color} is a notable attempt to match 130 basic colors and 1170 three-color combinations to 180 keywords.
Based on this, Solli~\shortcite{solli2010colrosem} proposes a semantic image descriptor that can be used as a tool in both image labeling and retrieval. 
Havasi~\shortcite{Havasi2010AutomatedCS} later generates single colors per meaning to benefit color selection (e.g. ``snow'' with  color ``white'').

\paragraph{Font \& typography.}  
Related studies of font and typography can be mainly summarized into the following two topics:
(1) \textbf{Recognition} aims to recognize font face from image. 
Features for classification can either be manually extracted, such as local typographical features, connected component and global texture analysis~\cite{sun2006fontregstroke}, or font embeddings learned using autoencoder~\cite{Wang2015DeepFont} as mentioned in Section~\ref{subsec:repr-element}. 
(2) \textbf{Structuring} organizes fonts by measuring their visual similarity instead of listing them in an alphabetical order.
 O'Donovan~\shortcite{odonovan2014font} estimates font similarity by training a model with predefined features such as italics and thickness.
Zhao~\shortcite{Zhao2018ModelingFI} learns a font face embedding space and train a multi-task DNN by predicting font properties.




\subsection{Design-level Understanding}

Now we focus on the overall analysis of a design.


\paragraph{Layout.} 
Layout analysis aims to parse the logical structure (i.e., object roles and their relations) of an unstructured design.
(1) \textbf{Object roles}: Most of previous works focus on scanned text documents such as academic articles and forms.
Detailed related works can be referred to Binmakhashen~\shortcite{galal2019survey}.
Compared to text documents with simple rectangle grid layouts, other types of designs include more complex layout variations with free-form and overlapped objects, which makes layout analysis more challenging and less explored.
Previous works mainly adapt the image detection/segmentation models and incorporate some unique characteristics of graphic design into modeling.
For example, as object edges provide skeleton information, E3Nets~\cite{wu2021e3net} embed the edge features along with the RGB channels to train a pixel segmentation model. 
Meanwhile, MagicLayouts~\cite{Manandhar2021magic} integrates the co-occurrence between UI components as a structural prior into the existing object detector.
Hao~\shortcite{hao2023reverse} further adopts the advanced Transformer-based detection architecture.
(2) \textbf{Relations}: some works further consider the spatial and semantic relation parsing in a design.
Rule-based methods~\cite{Xu2016WebGroup} incorporate the Gestalt laws of proximity to infer the elements groupings in HCI designs or web pages. 
As for learning-based methods, CanvasEmb~\cite{Xie2021CanvasEmb} first pretrains a large-scale design representation and then predicts the relation of an object pair with  finetuning.
Following the same architecture, Shi~\shortcite{shi2022reverse} recovers the hierarchical structure by recursively merge objects with an empirical threshold.

\paragraph{Overall design.} It studies the human perception of the overall design that can be a mixed result from different visual elements. Here we introduce two popular topics, visual style and visual flow. 
(1) \textbf{Visual style.} 
The overall look and feel of a design needs to cohere with its content. 
Style labels describing a design (e.g., artistic, dynamic, energetic, fresh) are selected from crowd-sourcing or from existing design books. 
To recognize the style, most of the previous works view it as image classification~\cite{huang2020visual}. 
Instead of binary classification objective, they learn a ranking function to estimate the score to a style with hand-crafted features~\cite{Chaudhuri2013AttribitCC} or automatically-learned neural features~\cite{zhao2018personality}. 
Furthermore, Zhao~\shortcite{zhao2018personality} conducts a detailed analysis on how various design factors affect style across different scales (from pixels, regions to objects). 
(2) \textbf{Visual flow.} It studies where the viewers look and how their visual flows move on a design, which can serve as a clue to guide designers to improve their works. A large amount of works have been built based on eye tracking data collection and analysis. 
They rely on hand-crafted features such as human face and positional bias, relative size and color contrast between object pairs~\cite{pang2016flow}. 
Recently deep learning based models~\cite{Zheng2018websaliency} generally encode the design as image using CNN and output the saliency map by predicting the pixel-level importance. 

\subsection{Discussion}
In this section, we try to generalize several key aspects in graphic design understanding. Previous works have considered many similarly-defined tasks but targeting for different scenarios, where most of the annotated datasets and proposed models can be hardly reused or transferred. We argue that the capability of generalization should be more considered in dataset construction and model design, and more unified frameworks are encouraged for solving these separate tasks more efficiently.
Furthermore, explainability would be useful in model output, especially for graphic design with complex compositions (e.g., model is expected to explain the association between the \textit{organic} style in Figure~\ref{fig:understanding} and the use of dominant green color).
\section{Generation}
\label{sec:gen}
Generation aims to automate the complicated creation process of graphic designs.
Existing studies investigated how to create an individual design element, i.e., \emph{element-level generation}, and how to compose different design elements to produce a complete design, i.e., \emph{design-level generation}.

\subsection{Element-level Generation}
As creating different elements requires different technologies, we categorize the studies in this line by the design element.

\paragraph{Shape.}
The shape is usually represented by Scalable Vector Graphics (SVG), which is a sequence of higher-level commands paired with numerical arguments.
We categorize existing approaches for shape generation by the supervision signal they utilize.
First, some work directly uses \emph{SVG path as supervision}.
Specifically, Lopes~\shortcite{Lopes2019Learned} trains a VAE on raster images to get the latent code and an autoregressive SVG decoder to predict SVG paths from the learned latent code; Carlier~\shortcite{carlier2020deepsvg} exploits the hierarchical nature of SVG in both the encoder and decoder of VAE by disentangling high-level shapes from the low-level commands.
Furthermore, since it is difficult to collect large-scale and high-quality SVG paths, Reddy~\shortcite{Reddy2021Im2Vec} proposes using \emph{raster image as supervision}.
It trains an encoder to map a raster image to a latent code which is then decoded into a SVG, and then uses a differentiable pipeline to render the generated SVG onto a raster image which is compared to a raster ground truth for training.
Moreover, as language-image encoder (e.g., CLIP~\cite{radford2021learning}) and text-to-image generation models (e.g., DALL-E~\cite{ramesh2022hierarchical} and Stable Diffusion~\cite{rombach2022high}) achieve great success, recent work considers leveraging \emph{pretrained model as supervision}.
Frans~\shortcite{frans2021clipdraw} optimizes the position and colors of the Bezier curves so that the generated drawings best match the given textual description, where a pretrained CLIP model is used as a metric for the similarity between the given description and the generated drawing.
Jain~\shortcite{jain2022vectorfusion} first rasterizes the SVG given path parameters and then back-propagates the score distillation loss computed from Stable Diffusion to update paths.

\paragraph{Color.} 
Mainstream approaches for color generation can be classified into two groups, i.e., \emph{palette-based methods} and \emph{object-based methods}.
Palette-based methods aim to generate a harmonic and aesthetic color palette, which is expressed by a limited number of colors in a fixed form.
Jahanian~\shortcite{jahanian2013automatic} produces the color palette by studying well-known color theories (including color harmony and color semantics) and quantifying them in a mathematical framework.
Instead of utilizing color theories, Qiu~\shortcite{qiu2022color} leverages a large-scale dataset of vector graphic documents to learn to generate the color palette.
It models color palettes as a sequence and proposes a masked color model based on BERT architecture~\cite{BERT} for color sequence completion.
After generating color palettes, palette-based methods rely on human effort or heuristic rules to match the colors in the color palette to the objects in the graphic design.
Differently, object-based methods directly generate colors for the objects.
Specifically, Yuan~\shortcite{yuan2021infocolorizer} frames color recommendation for infographics as a conditional generation problem, where the conditions include the observed colors of certain objects and non-color features (e.g., tree structure).
It employs Variational AutoEncoder with Arbitrary Conditioning (VAEAC), where which part is the conditions is controlled by a binary mask vector.

\paragraph{Font \& Typography.} 
To make the overall design aesthetically pleasing, the selection of fonts takes into account the existing information in a graphic design.
Choi~\shortcite{choi2018fontmatcher} considers the information of the images.
It recommends fonts that convey similar feelings to the given image (e.g., a warm feeling).
To achieve such a goal, it first utilizes an image impression model to map the image into a semantic space and then proposes a font impression model to calculate the distance between the given image and each font in the same semantic space.
Jiang~\shortcite{jiang2019visual} considers the information of the texts.
It recommends a font to be used in a part (e.g., body) given a font for another part of a document (e.g., header).
Specifically, the font recommendation is learned from large-scale human-generated font pairs by dual-space kNN and asymmetric similarity metric learning.

\subsection{Design-level Generation}
When design elements are composed to generate a complete design, lots of effort goes into the following tasks: 1) arriving at an appropriate \emph{layout}, i.e., the spatial relationships between objects, and 2) optimizing the \emph{overall design} choices, i.e., the comprehensive consideration of all the elements.

\paragraph{Layout.}
The layout is represented as positions and sizes of all the objects in a design.
We categorize layout generation tasks into three groups by the constraints they condition on. 

\emph{Unconditional generation} aims to generate high-quality and diverse layouts without considering any constraints.
Li~\shortcite{li2019layoutgan} proposes a novel generative adversarial network (GAN) where the generator uses self-attention modules to refine randomly-placed 2D graphic objects and the discriminator maps the generated layout to a wireframe image.
Patil~\shortcite{patil2020read} incorporates the hierarchy information of document layout into Variational Autoencoder (VAE). 
Arroyo~\shortcite{arroyo2021variational} proposes using Transformer~\cite{Transformer} as the building blocks of VAE to capture high-level relationships between objects.
Jiang~\shortcite{jiang2022coarse} proposes to decompose the decoding process of VAE into two stages, where the first stage predicts coarse region information and the second stage generates the detailed placement for each object.

\emph{Generation conditioned on object wireframe} pays extensive attention to modeling diverse constraints about object wireframes.
First, some work considers object relationships as the constraint (e.g., the title is on the top of the image)~\cite{yang2023intermediate,weng2023graph,guo2021layout}.
Specifically, Lee~\shortcite{lee2020neural} proposes using graph neural networks to model object relationships, and Kikuchi~\shortcite{kikuchi2021constrained} formulates the problem as a constrained optimization in the latent space of the model.
Second, Gupta~\shortcite{gupta2021layouttransformer} studies completion, which generates the placement of remaining objects given the placement of an initial set of objects.
It represents a layout as a sequence of graphical primitives and leverages Transformer decoder to predict the next primitive based on existing primitives in the layout.
Besides, Kong~\shortcite{kong2022blt} considers object types and sizes as the constraint.
It predicts the masked attributes by attending to surrounding attributes during training and refines the layout by masking out low-confident attributes iteratively during inference.
Moreover, Jiang~\shortcite{jiang2022unilayout} tackles various constraints by a uniform framework.
It represents constraints as sequences of tokens and leverages an identical encoder-decoder framework with Transformer to handle different constraints.

\emph{Generation conditioned on object content} focuses on how the object contents influence layouts.
First, Zheng~\shortcite{zheng2019content} considers the relationship between the layout and the topic, style, and purpose encoded in the visual content of the image and the meaning of the text.
It encodes multi-modal contents and the structural/categorical attributes of designs and leverages GAN to model layout distribution.
Besides, other work studies how to place objects over a natural image, where the objects are better not to overlap significant areas of the image (e.g., human faces) and the overall appearance of the generated design is harmonious.
Specifically, Zhang~\shortcite{zhang2020smarttext} generates candidate positions and sizes based on the salience map of the image, and then develops a scoring network to assess the aesthetic quality of the candidates; 
Zhou~\shortcite{zhou2022composition} proposes a domain alignment module with an inpainting network and a salience detection network to restore the natural image from a design, aiming at solving the data shortage problem of the training;  
and Cao~\shortcite{cao2022geometry} adopts cross attention to fuse the visual information of the image and designs a geometry alignment module to align the geometric information of the image and the layout representation.

\paragraph{Overall Design.}
First, Yang~\shortcite{yang2016automatic} explores a template-based approach.
It proposes a set of topic-dependent templates incorporating aesthetic design principles and develops a computational framework to automatically crop the image and optimize the typography and color under the constraints from the template.
Besides, some work investigates multi-stage generation-based approaches, where layout, font, and color are handled by different modules.
Specifically, Vaddamanu~\shortcite{vaddamanu2022harmonized} develops a genetic algorithm-based technology to generate layouts by considering visual salience, alignment and overlap, and then uses the generated layout to contextually select color and font; Jin~\shortcite{jin2022text2poster} generates layouts by cascaded auto-encoder and stylizes color and font by a matching-based method.
Moreover, Yamaguchi~\shortcite{yamaguchi2021canvasvae} proposes a single-stage generation-based approach, where the whole design is generated by a single model.
It defines a graphic design by a multi-modal set of attributes associated with a canvas and a sequence of visual objects such as shapes, images or texts, and trains VAE to learn the representation.

\subsection{Discussion}
In general, layout generation receives more attention compared to element-level generation and overall design generation, which are also critical for creating a pleasing design and worth of more exploration.
Besides, how to control the generation by user requirements is not well-studied.
On the one hand, some practical and friendly ways to express user requirements, e.g., text and sketch, do not receive enough attention.
On the other hand, it is worth of further investigation about how to reduce the violation of user constraints and improve the generation quality at the same time.
Furthermore, some important features for graphic design, which can make the generated design more attractive and creative, are not discussed by existing studies.
For example, for font, while font face and size have been studied, other critical attributes, e.g., line space, text alignment and font weight, are neglected; for color, gradient color and transparency are ignored.
Moreover, existing studies mainly focus on generating a single design.
In practice, a series of designs, e.g., a set of slides and posters, is also widely used.
To generate a series, the consistency between designs should be taken into account, which receives little attention currently.

\section{Future}
\label{sec:future}


Despite extensive efforts has been made, there remains many unresolved research problems. 
In this section, we summarize several promising directions to follow up.

\subsection{Benchmarks and Evaluation}
Although there are several benchmark datasets (see Table~\ref{tbl:dataset}), it is still an urgent need to build a better dataset due to two reasons.
The first one is the \emph{scale}.
The largest dataset about design intelligence contains 330k data, which is much smaller than the datasets in CV and NLP areas.
We believe a sufficient amount of data will promote the development of the algorithms.
The second one is the \emph{coverage}.
Current datasets usually support one or two tasks on a specific type of graphic design.
We think a better dataset that supports as many as tasks and contains various graphic designs will significantly benefit the learning and analysis of the algorithms.
Besides datasets, evaluation metrics are also critical.
To make the evaluation more comprehensive, we think proposing new metrics based on typical design principles is promising. 
Furthermore, as aesthetics is subjective, large-scale and formal human evaluation should be paid more attention to.



\newcommand{\tabincell}[2]{
\begin{tabular}{@{}#1@{}}#2\end{tabular}
}

\begin{table*}
  \begin{center}
    \begin{small}
        \begin{tabular}{llll}
            \toprule
            Dataset & Size & Tasks & Annotation \\
            \midrule
            DSSE & 200 & Layout analysis & Semantic role labels \\
            VisMet & 350 & Shape metaphor & Online ads images with metaphor keywords \\
            Image ads & 500 & Layout generation & Image banner ads with layout annotation \\
            AdobeVF & 1k & Font recognition & Text images with font class \\
            CTXFont & 1k & Font recognition & Web designs with font class \\
            Imp-1k & 1k & Visual attention & Pixel-wise importance maps \\
            Magazine & 4k & Layout generation & Images of magazine page \\
            Poster & 5k & Style classification
            & Online posters with style labels \\
            PAT & 10k & Color understanding & Palette-text pairs \\
            InfoVIF & 13k & Color generation & Infographics obtained from design resources sites \\
            Crello & 23k & Color generation, layout generation & Designs with canvas and elements in vector format \\
            InfoPPT & 23k & Layout analysis & Slides with hierarchy \\
            RICO & 91k & Layout analysis, layout generation & UI designs collected from mobile apps \\
            SVG-Icons8 & 100k & Shape generation & SVG icon shapes \\
            PubLayNet & 330k & Layout analysis, layout generation & Images of papers and annotations for elements \\
            SVG-Fonts & 14M & Shape generation, font recognition & Fonts across 62 characters in SFD format \\
            QuickDraw & 50M & Shape recognition, shape generation & User sketches with stroke action and shape class \\
            \bottomrule
        \end{tabular}
    \end{small}
    \caption{Benchmark datasets related to graphic design.}
    \label{tbl:dataset}
  \end{center}
\end{table*}


\subsection{Large-scale Self-supervised Learning}
Large-scale self-supervised learning has mede incredible progress in natural language and computer vision.
It aims to learn salient representations from a large amount of unlabeled data, where downstream tasks can leverage the learned features for better performance.
While preliminary attempts~\cite{Xie2021CanvasEmb,UIBert,He2021ActionBert} have been made for graphic design along this direction, there are still many challenges and opportunities.


\paragraph{Information Extraction and Fusion.}
A design can be viewed as either a pixel image or structured document with object metadata (e.g., type, hierarchy).
Due to such characteristics, there are several important questions when performing large-scale self-supervised learning for graphic design.
First, as designs are usually stored as pixel images on the web, it is not easy to transform it into a structured document.
Recently, reverse engineering for graphic design~\cite{shi2022reverse}, which recognizes primitive objects from design images, has been attracting growing interest.
It will be a promising direction to obtain rich structured design data. 
Second, whether all the information is critical and necessary.
Most of the existing approaches only consider a subset of metadata while there is little discussion about the informativeness of different metadata.
Third, how to effectively fuse information in various modalities. It is worth exploring the combination of design in both structure and pixel formats.

\paragraph{Model and Pretext Tasks.}
Existing approaches similarly adopt classical Transformer as the backbone and use masked language modeling (MLM) as the pretext task.
We expect more discussion on different model architectures and pretext tasks.
For example, CNN-based architecture or vision Transformer could be a good choice.
For MLM, it would be better if there is a thorough discussion about which part of graphic design should be masked.
Besides MLM, the completion mechanism, which is widely used in natural language~\cite{brown2020language}, is also worth to be explored for graphic design.

\subsection{Design Knowledge Formulation and Encoding}
Design knowledge, commonly in the form of design principles, serves as the foundation in graphic design.
It is a good reference and standard for interpreting the design logic (e.g. where to place the object for visual attraction, what color to use for highlighting the contrast) and evaluating the quality.
Furthermore, it can be used as an indicative feature in downstream tasks to represent, analyze, and generate designs.
There are several attempts to explicitly express the principles into machine learning models~\cite{kikuchi2021constrained,Manandhar2021magic,Xie2021CanvasEmb}, but they are limited to some principles that are easy to measure, such as alignment and overlap.
Hence, there is still a large exploration space of better formulating and encoding more design knowledge.

\subsection{Human-AI Collaboration}

AI has enhanced many applications in graphic design~\cite{li2021towards,evan2020humanoutperform} and raised lots of discussions about its positive and negative impacts. 
As graphic design is human-centered, it is important to emphasize the role of AI as an effective assistant. AI should but not be limited to follow human-centered design principles, respect human’s cognitive capacities and be responsible to ethics.
In each step of the design process including material searching, ideation, creation and refinement, how AI can optimally intervene and collaborate with human requires lots of user studies and experiments.
So far the topic of human-AI collaboration is less explored. 
We hope more research and development of human-AI collaboration will be conducted in the future to create an effective and harmonious design environment for human with maximizing the value of AI.

\section{Conclusion}

This paper presents the first comprehensive survey in the field of graphic design intelligence. We summarize relevant AI techniques with respect to representation, understanding, and generation. In each direction, we consider both the visual-element-level and overall design-level related works.
Moreover, we shed light on several promising future directions. We hope this survey can provide a clear overview of graphic design intelligence, and inspire researchers with new discoveries to advance the development of this field.

\bibliographystyle{named}
{
\small
\bibliography{ijcai23}
}

\end{document}